\documentclass[prb,twocolumn,showpacs]{revtex4}
\usepackage{graphicx}
\usepackage{graphics}
\usepackage{dcolumn}
\usepackage{bm}
\newcommand{\beq}{\begin{equation}}
\newcommand{\eeq}{\end{equation}}
\newcommand{\beqnar}{\begin{eqnarray}}
\newcommand{\eeqnar}{\end{eqnarray}}
\newcommand{\bfig}{\begin{figure}}
\newcommand{\efig}{\end{figure}}

\begin{document}
\title{Edge proximity-induced magnetoresistance and spin polarization in ferromagnetic gated bilayer graphene nanoribbon}

\author{Vahid Derakhshan, Hosein Cheraghchi}

\affiliation{School of Physics, Damghan University, P. O. Box:
36716-41167, Damghan, IRAN } \email{cheraghchi@du.ac.ir}
\date{\today}

\begin{abstract}
Coherent spin-dependent transport through a junction containing
of Normal/Ferromagnetic/Normal bilayer graphene nanoribbon with
zigzag edges is investigated by using Landauer formalism. In a
more realistic set-up, the exchange field is induced by two
ferromagnetic insulator strips deposited on the ribbon edges
while a perpendicular electric field is applied by the top gated
electrodes. Our results show that, for antiparallel
configuration, a band gap is opened giving rise a semiconducting
behavior, while for parallel configuration, the band structure
has no band gap. As a result, a giant magnetoresistance is
achievable by changing the alignment of induced magnetization.
Application of a perpendicular electric field on the parallel
configuration, results in a spin field-effect transistor where a
fully spin polarization occurs around the Dirac point. To be
comparable our results with the one for monolayer graphene, we
demonstrate that the reflection symmetry and so the parity
conservation fails in bilayer graphene nanoribbons with the
zigzag edges.
\end{abstract}
\pacs{Spin polarization, magnetoresistance, ferromagnetic bilayer
graphene, perpendicular applied gate voltage} \keywords{}
\maketitle
\section{Introduction}
\bfig
\includegraphics[width=8cm,height=8cm,angle=-90]{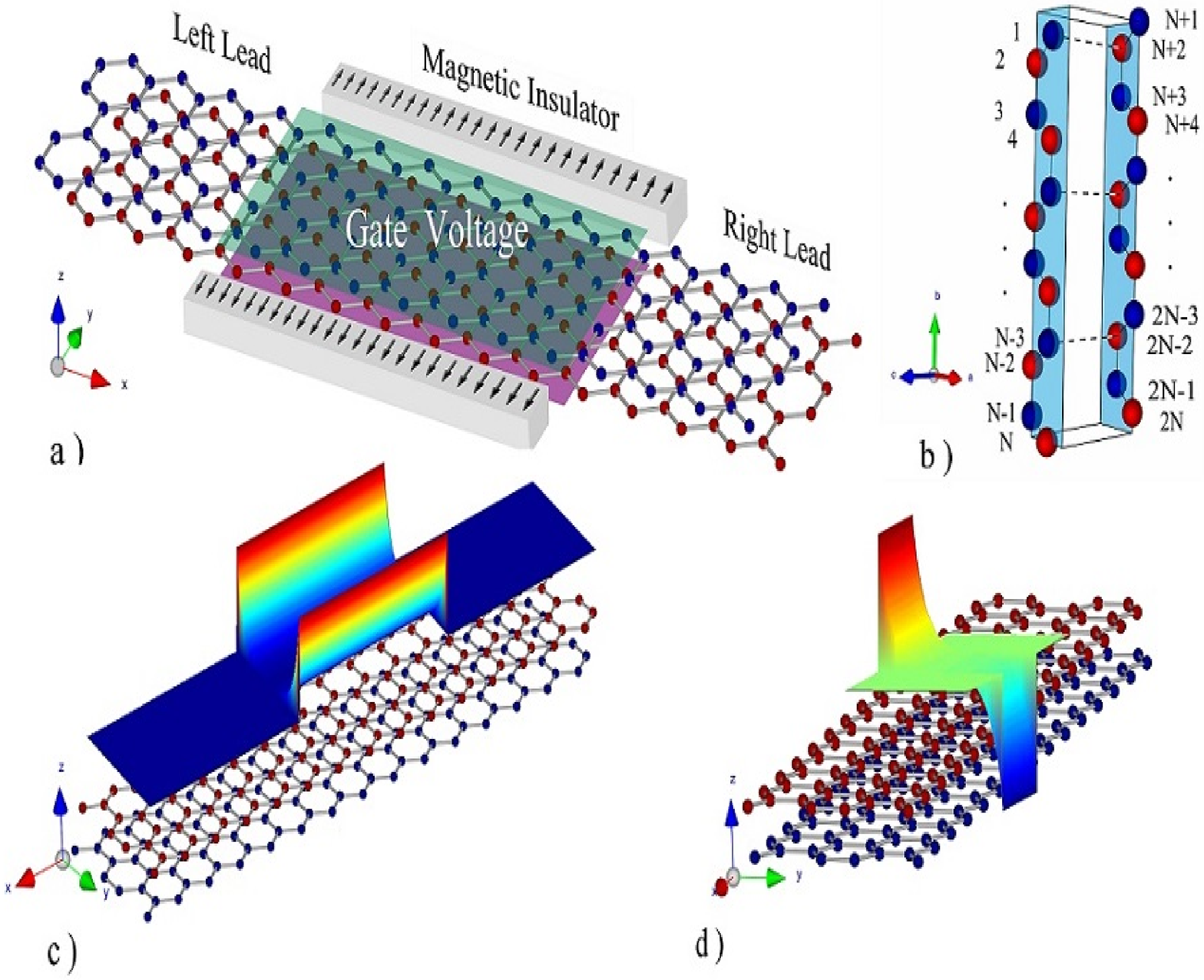}
\caption{ a)Schematic geometry of two-terminal structure for
zigzag bilayer graphene nanoribbon (ZBGN) in the $AB$ Bernal
stacking. b) Unit cell of the central region and the left and
right electrodes. The unit cell consists of $2N$ atoms. Two
ferromagnetic strips are attached to the edges of nanoribbon
which induce exchange field in ZBGN. The direction of
magnetization of strips can be parallel or antiparallel along the
direction perpendicular to the graphene sheet. The induced
magnetic potential $M$ exponentially decays from the ribbon edges
toward its middle. The induced magnetic field is arranged as (c)
the parallel configuration or (d) the anti-parallel configuration
of the edge magnetization. The decaying length of the induced
magnetic potential is $1/3$ of the ribbon's width.} \label{fig1}
\efig
The recent successful fabrication of monolayer
graphene\cite{novoselov} has been attracted great attention
because it prepares appropriate building block for future
nanoelectronics and spintronics. The Dirac Hamiltonian at low
energy limit of graphene spectrum results in many peculiar
properties, for instance, the half integer quantum Hall
effect\cite{qhe}, minimum conductivity\cite{mincond}, the absence
of back scattering\cite{mincond,back-scatter}, Klein
tunneling\cite{klein}. Furthermore, graphene has excellent
transport properties such as long spin relaxation which makes it
excellent candidate for spintronic devices. Because of weak spin
orbit coupling, spin relaxation length of graphene can reach
about one micron in dirty samples and room temperature
\cite{tombros}. Clean samples have longer spin coherency length.
Carriers in graphene are not spin-polarized because graphene has
not intrinsically ferromagnetism (FM) properties. However, FM can
be induced extrinsically in graphene by doping
defects\cite{defects}, Coulomb interactions\cite{coulomb} or by
applying an external electric field in the transverse direction
in nanoribbons\cite{electric}. One of proposals\cite{Haugen,joza}
is the proximity effect of a FM insulator deposited on graphene
sheet. Because of strong proximity, the wave functions of the
localized magnetic states overlap with carriers in graphene sheet
giving rise an induced exchange field in graphene. This induced
exchange field which is tunable by application of an in-plane
external electric field \cite{ex-tune} induces a spin-polarized
current in graphene. Using a ferromagnetic gate dielectric which
is deposited on graphene layer as the channel, a spin field
effect transistor has been proposed when spin manipulation is
achievable by applying external perpendicular electric
field\cite{semenovAPL}. External electric field can modify the
exchange interaction.

The possibility of controlling spin conductance in monolayer
graphene has also been recently studied\cite{Yokoyama}. It was
shown that because of induced exchange field and also due to the
chiral resonant tunneling bound states inside the barrier, spin
splitting of current emerges and has an oscillatory behavior in
respect to the gate voltage and chemical potential. On the other
hand, transport gap induced by the parity selection rule
governing in the electronic bands of spectrum of zigzag graphene
nanoribbons leads to large spin polarization and giant
magnetoresistance\cite{zhang}. Presence of the reflection
symmetry in the incoming and outgoing wave functions of zigzag
monolayer graphene nanoribbons leads to the parity selection rule
which regulates the current flow\cite{cheraghchi}. We will show
in this paper that there is no such a symmetry and selection rule
in bilayer graphene. Experimentally, possibility of fabrication
graphene nanoribbons with ultra narrow widths and atomically
smooth edges possibly well-defined zigzag or armchair-edge
structures has been recently reported\cite{Dai}. In zigzag
graphene nanoribbons, spin current has been also predicted in
presence of large electric field\cite{electric}. Large
magnetoresistance (MR) also has been reported in monolayer
graphene nanoribbons\cite{kim}.

Much attention has been recently paid to bilayer graphene which
consist of two parallel graphene sheets coupled each other with
two sublattices $A$ and $B$, in each layer. They are typically
stacked in $AB$ Bernal form as shown in Fig.\ref{fig1} leading to
some interesting physical phenomena. For example new type of
quantum Hall effect\cite{qhe-bilayer} and also the energy band
gap tunable by vertically applied electric field are of its
peculiar properties in compared to monolayer
graphene\cite{4band,gap-BG,gap-BG1,gap-BG2}. A $200$ meV band gap
for bilayer graphene has been proposed by optical measurements
and also theoretical predictions. Recently, an electrically
tunable band gap has been observed in trilayer graphene with
$ABC$ crystallographic stacking\cite{trilayer}. Controllable band
gap introduces bilayer graphene as an excellent candidate for
spintronic devices\cite{semenov}. The same as monolayer graphene,
in the resonant bound states of the barrier, large spin
polarization and also giant magnetoresistance has been reported
in bilayer graphene in the limit of infinite
width\cite{nguyen,Yu,adineh}. Motivated by the above mentioned
studies we consider spin polarization and magnetoresistance in
bilayer graphene nanoribbon with zigzag edges.

Mostly, FM insulator layers are deposited on top or bottom of
BLG. However, simultaneous application of a perpendicular electric
field which can be manipulated by means of the top-gate
electrodes, affects the induced exchange field arising from a FM
insulator layer. In this work, the exchange field is induced by
the FM insulator strips deposited on the ribbon edges while a
perpendicular electric field is applied by means of the top-gate
electrodes. This structure is more realistic from the
experimental point of view. So we have to deposit the gate
electrodes separate of the FM insulator strips. A schematic
cartoon of our considered system is shown in Fig.\ref{fig1} which
contains the $AB$ stacking of bilayer graphene nanoribbon
accompanied with two strips of ferromagnetic insulator such as
$EuO$ deposited on two edges of nanoribbon. The direction of
magnetization of the strips can be essentially justified in the
parallel or antiparallel configurations. The induced exchange
field is directed to perpendicular of graphene sheet. To simulate
more realistic situation, we suppose that the edge-induced
magnetic potential exponentially decays from the edge toward the
middle of nanoribbon. In the tight-binding model, we use
Landauer-Buttiker formalism to calculate the spin-dependent
conductance. We have found large spin polarization and
magnetoresistance when a band gap is opened in the antiparallel
configuration. The energy range in which spin polarization and
magnetoresistance are large, can be extended by application of a
perpendicular gate voltage. Based on this behavior, one can
manipulate a spin filed effect transistor.

The paper is organized as the following sections: in section
~\ref{Sec:TheProposedMethod}, we introduce the model and
Landauer-Buttiker formalism for calculating conductance. The
numerical results and discussion about spin dependent conductance
and spin filtering and also magnetoresistance are presented in
section ~\ref{Sec:numericalresults}. violation of parity
conservation in bilayer graphene nanoribbons with even number of
zigzag chains in width is presented in section
~\ref{Sec:ref-symmetry}. Finally a conclusion of contents is
presented in section ~\ref{Sec:conclusion}.

\section{Hamiltonian and Formalism}\label{Sec:TheProposedMethod}
We consider a bilayer graphene nanoribbon with Bernal stacking
$(AB)$ which is connected to the left and right electrodes as
indicated in Fig.\ref{fig1}. The total Hamiltonian, $\textbf{H}$,
of the device can be divided into four parts as $ H_C, H_L, H_R,
H_T$ which are the Hamiltonian of the center region, the left and
right electrodes, and also the coupling Hamiltonian, respectively.
The electrodes are considered to be the same as the central
portion. The model is the tight-binding nearest-neighboring
approximation with one $\pi$ orbital per each site on the lattice.
The effective Hamiltonian of bilayer graphene in the presence of
magnetic strips is given as follows:

\begin{eqnarray}
H_{C,\sigma}=\sum_{l,i}(\varepsilon_{l,i}
+\lambda_{\sigma}M_{l,i})
a_{l,i,\sigma}^{\dagger}a_{l,i,\sigma}\nonumber
\\+\sum_{l,i}(\varepsilon_{l,i}+\lambda_{\sigma}M_{l,i})
b_{l,i,\sigma}^{\dagger}b_{l,i,\sigma} \nonumber \\
-t\sum_{l,\langle
i,j\rangle}(a_{l,i,\sigma}^{\dagger}b_{l,j,\sigma}
+H.C.)\nonumber \\ -t_{\perp}\sum_{\langle
i,j\rangle}(a_{1,i,\sigma}^{\dagger}b_{2,j,\sigma} +H.C.)
\end{eqnarray}

\begin{eqnarray}
H_{\alpha=L,R}=\sum_{l,i}
\epsilon_{l,i,\sigma}(a_{l,i,\sigma}a^{\dagger}_{l,i,\sigma}+b_{l,i,\sigma}b^{\dagger}_{l,i,\sigma})\nonumber
\\-t\sum_{l,\langle
i,j\rangle}(a_{l,i,\sigma}^{\dagger}b_{l,j,\sigma}
+H.C.)\nonumber \\ -t_{\perp}\sum_{\langle i,j
\rangle}(a_{1,i,\sigma}^{\dagger}b_{2,j,\sigma} +H.C.)
\end{eqnarray}
\\
where $(a_{l,i,\sigma}^{\dagger})$ and $(a_{l,i,\sigma})$ and
$(b_{l,i,\sigma}^{\dagger})$ and $(b_{l,i,\sigma})$ are creation
and annihilation operators of an electron with spin
$\sigma(\sigma=\uparrow,\downarrow)$ in the sublattice $ A(B) $
in the layer $ l=1,2 $ at the $ i $th site respectively. The
onsite energy is indicated by $\epsilon_{l,i}$. Here, $
\lambda_{\sigma}=\pm 1 $ for $ \sigma=\uparrow,\downarrow $. The
interlayer  nearest neighbor hopping energy is $ t=3.16 eV $
while the hopping energy between sublattices $ A $ and $ B $ in
different layers is $t_{\perp}=0.39
eV$\cite{prependicular-hopping}. The top and bottom gate voltages
are applied through a uniform variation in onsite energies
$\varepsilon_{l,i}$ belonging to the topper ($l=1$) and lower
($l=2$) layers.  As shown in Fig.\ref{fig1}, FM strips deposited
on the edges of nanoribbon induce the exchange field across BLG.
The overlapping of the wave functions of the localized magnetic
states existing in the FM strips with itinerant carriers of
graphene reduces strongly with the distance. So it is reasonable
to suppose an exponential decay for such overlapping with
distance. For antiparallel configuration of FM strips, the
induced magnetic potential exponentially decreases from the
values of $\textbf{M}$ and $ \textbf{-M} $ at the edges to zero
in the middle of the ribbon. For parallel configuration, the
induced magnetic potential reaches to its maximum value at the
edges of nanoribbon as $\textbf{M}$ and exponentially decays when
one goes from the zigzag edges toward the middle of the ribbon..
\begin{figure}
\includegraphics[width=9 cm,height=8cm]{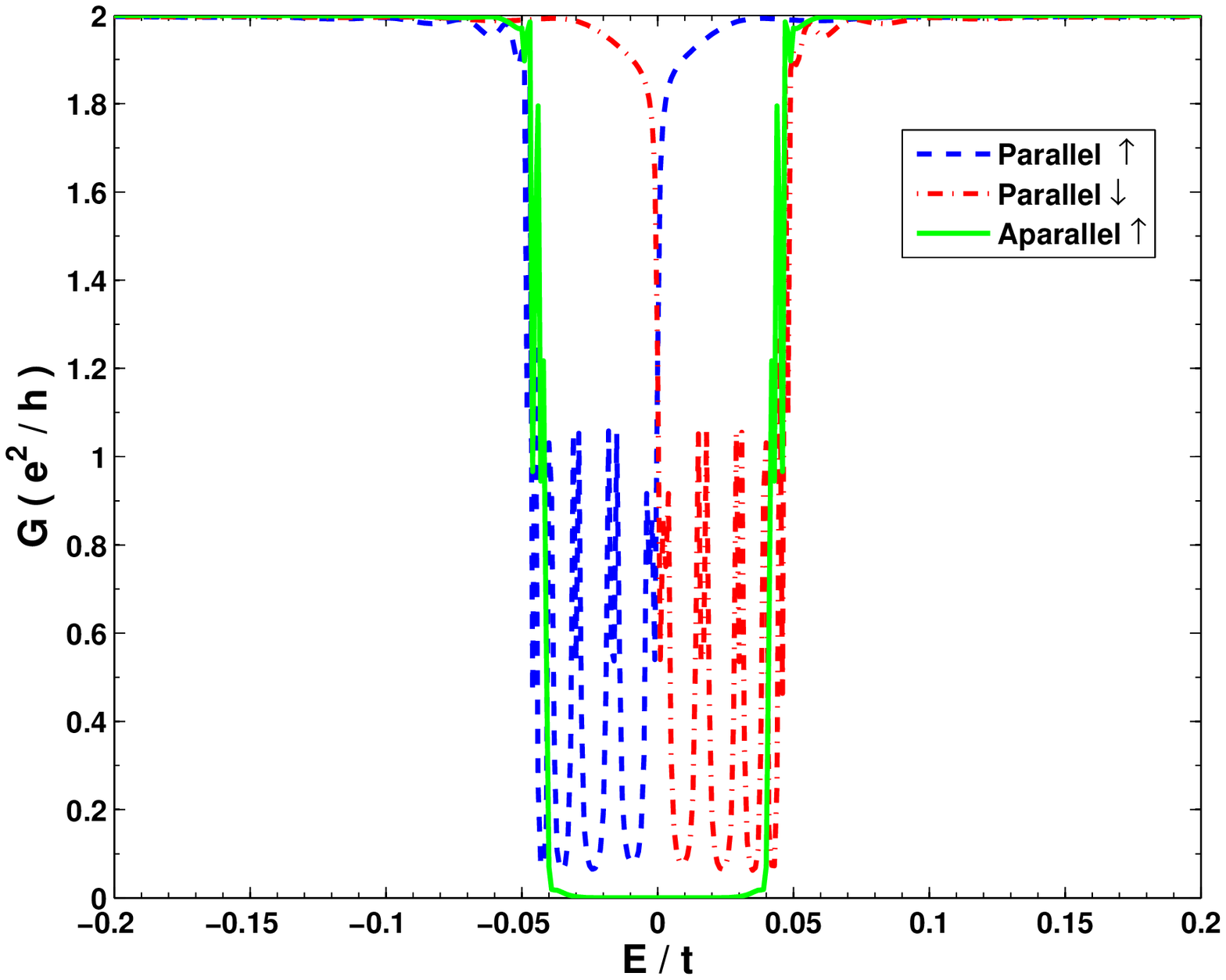}
\caption{Conductance versus dimensionless parameter $ E_F/t $ for
antiparallel (solid line) and parallel (dashed line for spin up
and dash-dotted line for spin down) with the induced magnetic
potential about $M=0.05t$.} \label{fig3}
\end{figure}
\begin{figure}
\includegraphics[width=9 cm,height=8cm]{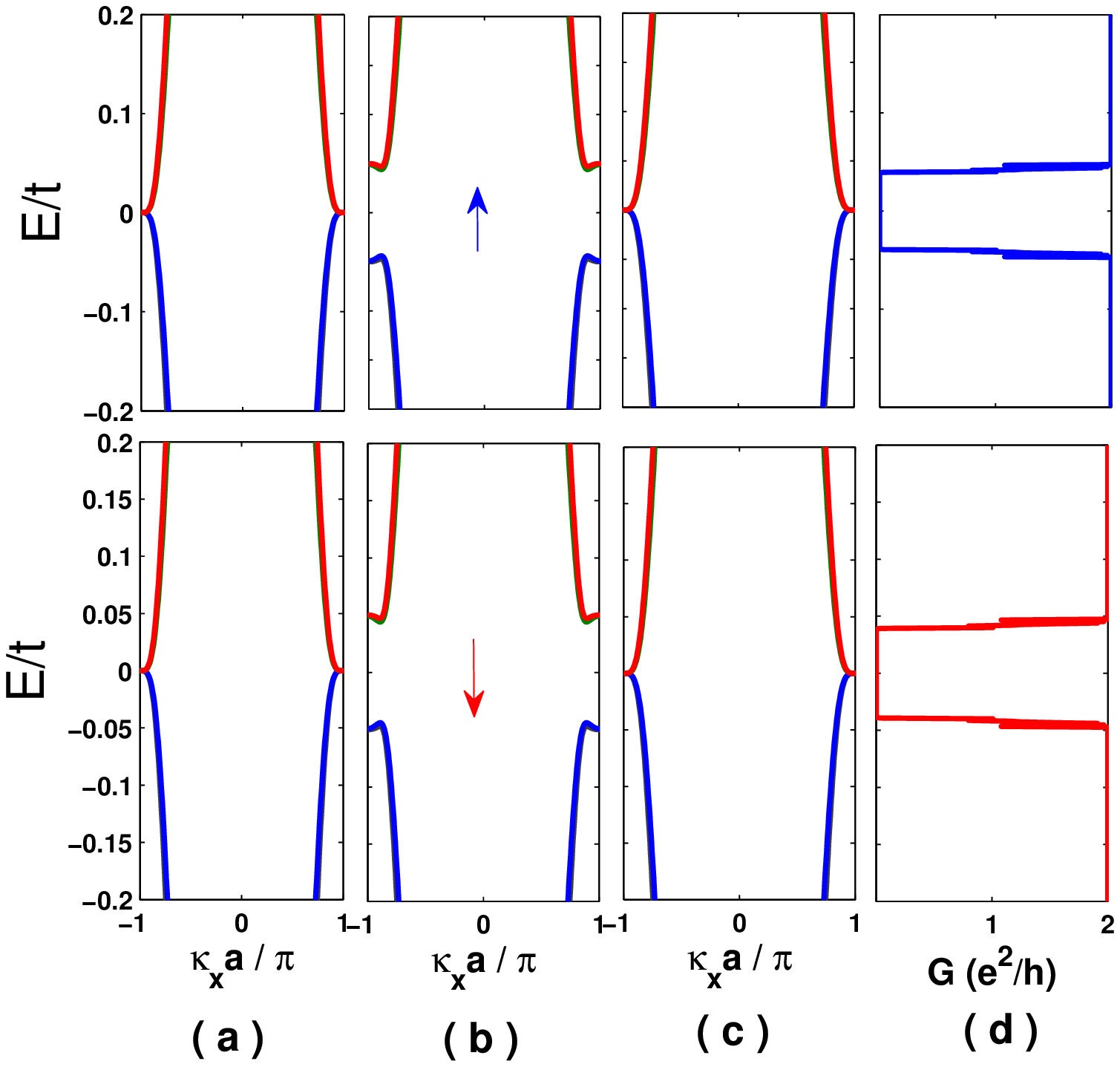}
\caption{Energy band structure of the left electrode, central portion and right electrode for antiparallel configuration of
magnetization of ferromagnetic insulator strips. A tunable band gap is opened in the band structure of the central portion in which
conductance exhibits a transport gap around the zero energy. Conductance as a function of Fermi energy for (a) spin up (b) spin down.}
\label{fig4}
\end{figure}

The current flowing through the device can be calculated from the Landauer-Buttiker formula,
\begin{equation}
I_{\sigma}=\frac{e}{h}\int dE T^{\sigma}_{LR}(E)[f_{L}(E)-f_{R}(E)]
\end{equation}\\
Where $f_{L/R}$ is the Fermi-Dirac distribution function of the
left and right electrodes and $
T^{\sigma}_{LR}(\epsilon)=Tr[\Gamma^{\sigma}_{L}G_{\sigma}^{r}\Gamma^{\sigma}_{R}G_{\sigma}^{a}]$
is the transmission coefficient. In the transmission calculation,
$ \Gamma_{\alpha}(\epsilon)$ is the rate of escaping carriers from
the central portion into the electrodes and is given by
$\Gamma^{\sigma}_{\alpha}=i[\sum^{r}_{\alpha,\sigma}(\epsilon)-\sum^{a}_{\alpha,\sigma}(\epsilon)]$
in which the related Green's function is written as $
G_{\sigma}^{r}(z)=[G_{\sigma}^{a}(\epsilon)]^{\dagger}=[z-H_{c,\sigma}-\sum^{r}_{L,\sigma}-\sum^{r}_{R,\sigma}]^{-1};
z=\epsilon+i \eta$ where $\eta$ is a positive infinitesimal
number. Here $\sum_{L/R}^{r} $ is the retarded self-energy
function emerging from semi-infinite electrodes. By calculating
surface Green's function \cite{nardelli} and coupling Hamiltonian
$H_T$, one can calculate self-energies related to each electrode.
After calculating the current, the conductance $G_{\sigma}$ can be
derived straightforwardly, $G_{\sigma}=lim_{V\rightarrow 0}
dI_{\sigma}/dV $, with the bias $ V=\mu_{L}-\mu_{R} $ where $\mu$
is the chemical potential of electrodes. At zero temperature and
linear regime, $ G_{\sigma}=\frac{e^2}{h}T^{\sigma}(E_F)
$\cite{Datta}.

\begin{figure}
\includegraphics[width=9 cm,height=8cm]{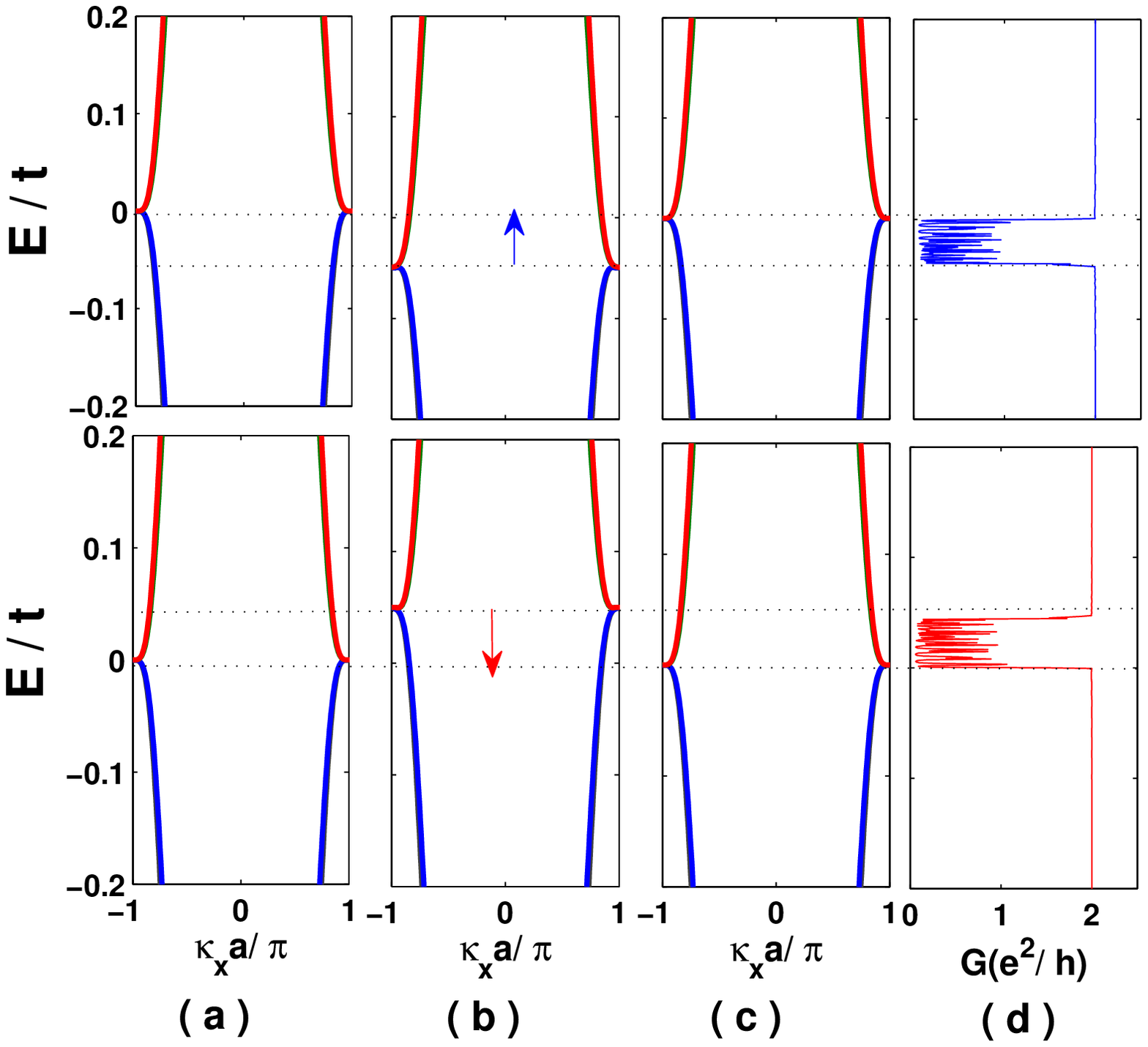}
\caption{Energy band structure of (a) the left electrode, (b) central portion and
(c) right electrode in the parallel configuration of magnetization of strips.
(d) Conductance as a function of Fermi energy. Topper (lower) curves are concerned to spin-up (down). Number of zigzag chains in width is equal to $15$ and number of unit cells in length is considered to be $100$.} \label{fig5}
\end{figure}

\begin{figure}
\includegraphics[width=9 cm,height=8cm]{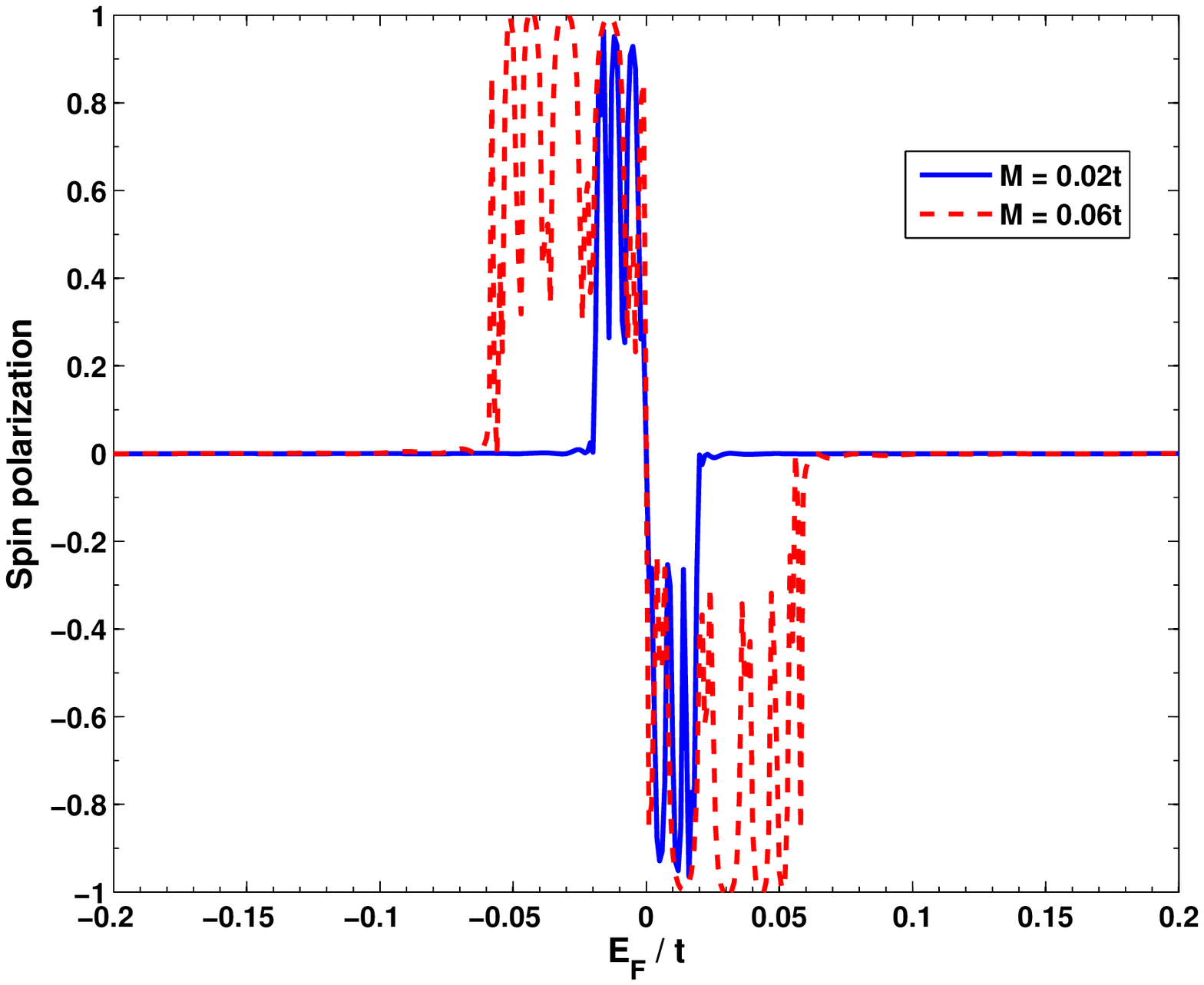}
\caption{Spin polarization as a function of Fermi energy for parallel configuration of magnetization of strips.} \label{fig6}
\end{figure}

\section{Numerical Results}\label{Sec:numericalresults}
{\it Conductance}: Let us start describing transport mechanism
for parallel and antiparallel configuration of magnetization of
strips. Fig.\ref{fig3} shows conductance as a function of the
Fermi energy for different configurations of FM strips. In the
parallel (antiparallel) configuration, the exchange field
inducing by FM insulators located on each edges are parallel
(antiparallel) and exponentially reduces from the zigzag edges to
the middle of the strip. Conductance for antiparallel
configuration shows a transport gap around the Dirac point which
is proportional to the spin splitting of the potential
originating from the induced exchange field. In fact, those
carriers with spin parallel to the induced exchange field (spin
up) hits on the barrier whose magnetic potential is lower than the
potential belonging to the spin down. For the parallel
configuration of FM strips, depending on the type of spin,
conductance oscillates with the amplitude about $\frac{2e^2}{h}$
in an energy range which is equal to the potential splitting.
This spin splitting of the potential barrier in the parallel
configuration leads to a large spin polarization around the zero
energy. By application of an electric field, one can control the
direction of the exchange field inducing by FM strips. So based on
this structure, observation of a large magnetoresistance around
the Dirac point is potentially accessible when parallel
configuration switches to antiparallel one by application of an
electric field.

\begin{figure}
\includegraphics[width=9 cm,height=8cm]{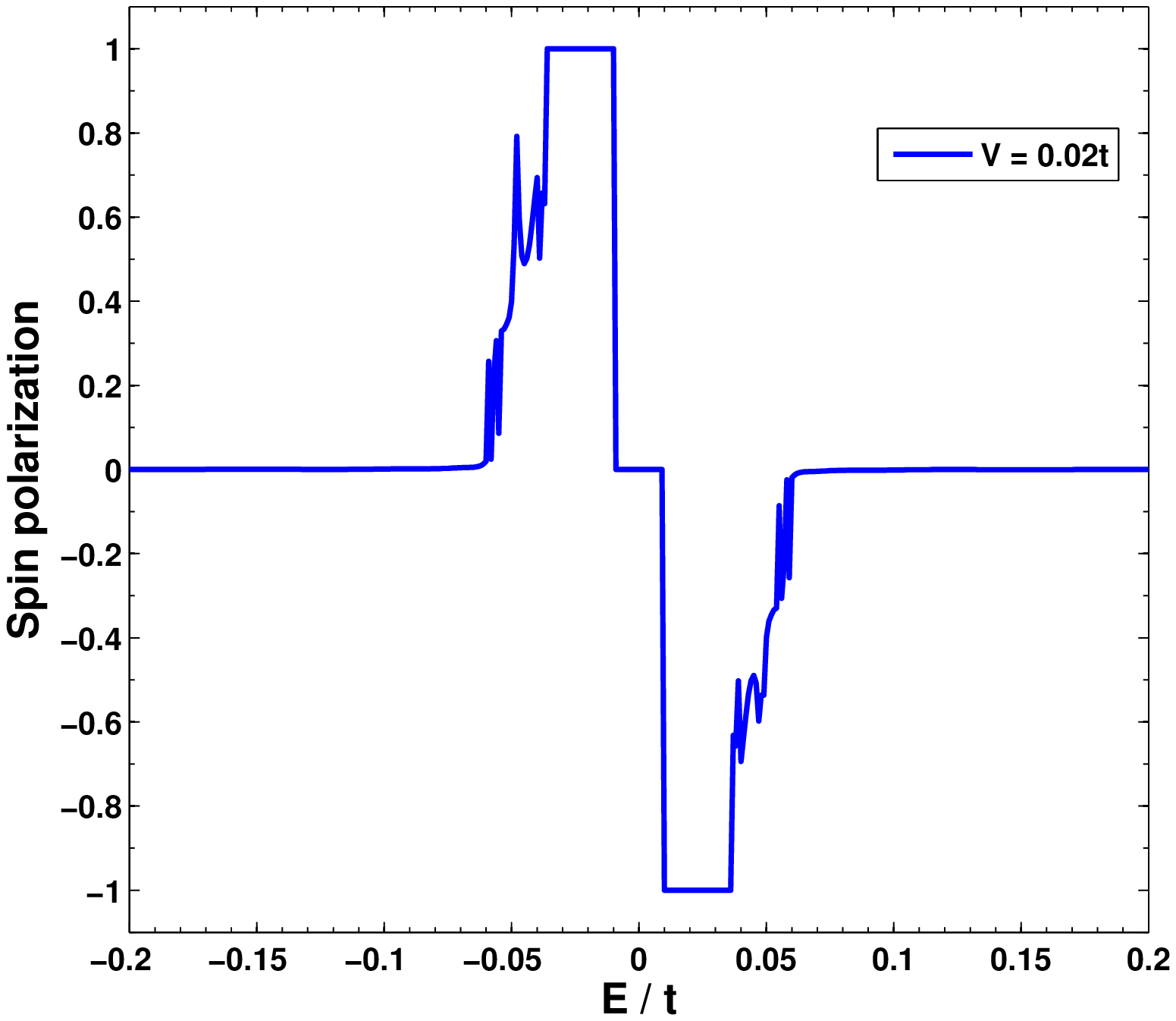}
\caption{Spin polarization as a function of Fermi energy for the
parallel configuration of magnetization of strips in the presence
of external perpendicular electric field. Potential difference
between the upper and lower layers is $V=0.02t$. The induced
exchange potential is equal to $M=0.05t$.} \label{fig7}
\end{figure}

To understand the zero conductance plateau seen in the
antiparallel configuration of FM strips, we plot the band
structure of the left electrode, central portion and right
electrode in Fig.\ref{fig4}. In the antiparallel configuration, a
band gap is opened in the band structure of the central portion
around the zero energy. The band gap is originated from breaking
of the chiral symmetry by application of an antiparallel exchange
field on two edge sides of bilayer nanoribbon. So in the
antiparallel configuration, system behaves as a semiconductor.
However, there is no energy gap in the band structure of the
parallel configuration. As shown in Fig.\ref{fig5}, in the
parallel configuration, spin-up (down) band structure of the
central portion shifts to lower (higher) energies as much as the
induced exchange potential. However, in the antiparallel
configuration, the band structure is independent of spin type.
Conductance in Fig.\ref{fig5} shows an oscillatory behavior which
is concerned to resonant states arising from induced magnetic
barrier shown in Fig.\ref{fig1}-d. We also observed that the
results are independent of even-odd effects of the number of
zigzag chains in width.

{\it Spin polarization}: In the parallel configuration, spin
splitting of conductance leads to large spin polarization around
the zero energy. The spin polarization is defined as:
\begin{equation}
P=\frac{G_{up}-G_{down}}{G_{up}+G_{down}} \label{polarization}
\end{equation}
where $G_{up}$ and $G_{down}$ are conductance for up and down
spins. Spin polarization in the parallel configuration is
indicated in Fig.\ref{fig6} for different induced exchange
potentials. In the energy range of $[0,M]$, there is a filter of
up spins, while in the energy range of $[-M,0]$, down spins are
filtered. So the range of energies in which spin filter is
observed increases with an increase of the induced exchange
potential.
\begin{figure}
\includegraphics[width=9 cm,height=8cm]{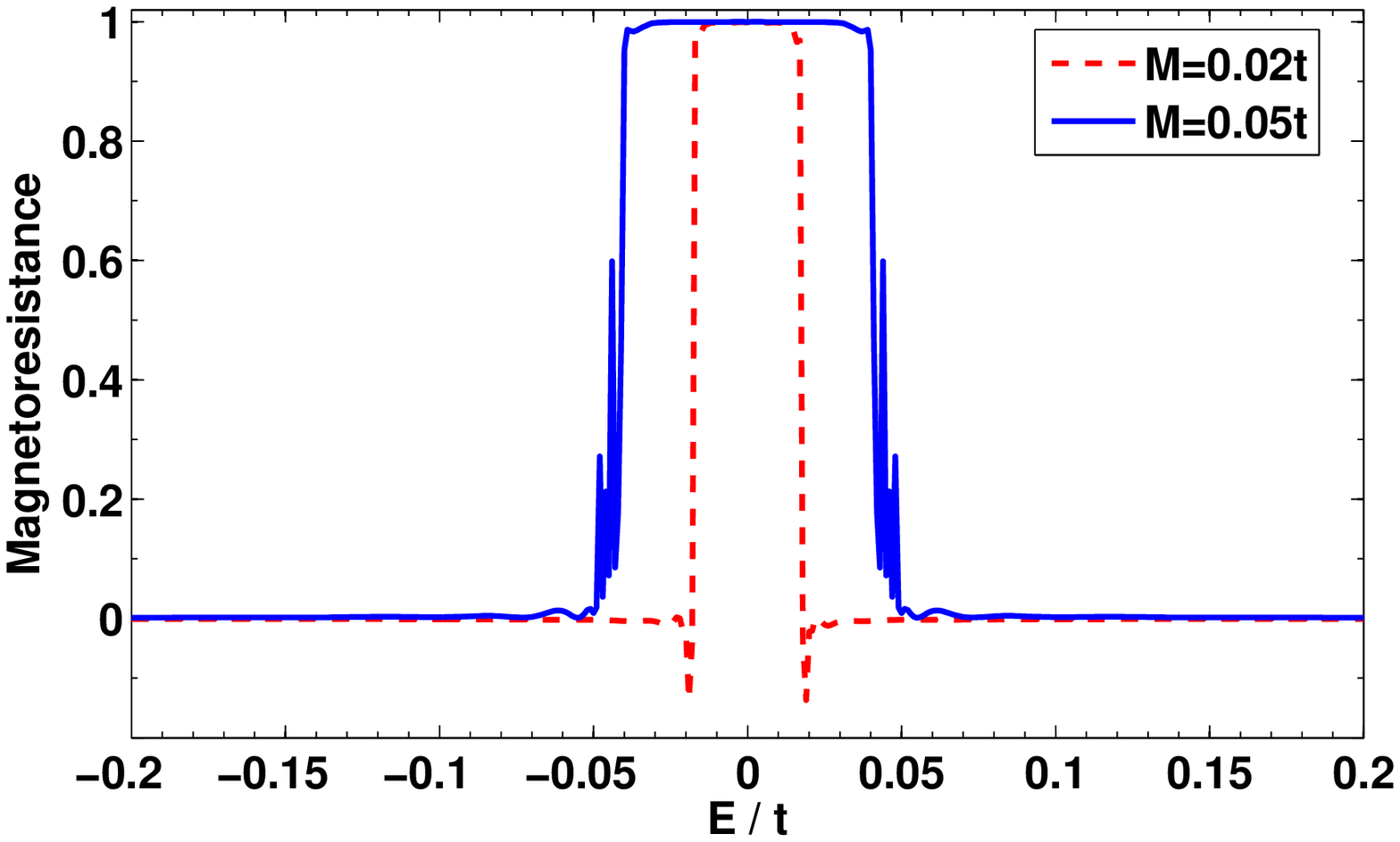}
\caption{Magnetoresistance as a function of Fermi Energy for
different values of the induced exchange field ($M$). Number of
zigzag chains in width is $6$ and number of unit cells in length
is $40$.} \label{fig8}
\end{figure}

The advantage of bilayer graphene against monolayer is the
existence of a controllable band gap which is produced by
application of a perpendicular electric field. In the case of
parallel configuration, Fig.\ref{fig7} indicates spin
polarization through ZBGN with $6$ zigzag chains in width and
$30$ unit cells in length in the presence of the external
perpendicular electric field $V$. Application of the potential
difference ($V=0.02t$) in the central region causes to emerge an
energy gap which is equal to $V$. So the first conduction and the
last valance bands detach from each other so that they do not
belong to the same band group. Therefore, as shown in
Ref.\onlinecite{cheraghchi-book}, transition between these
subbands is not allowed now. It means that detaching of bands
causes to block electron transition in the energy ranges of
$[V,V+M]$ and $[-V,-V-M]$. So two full spin polarized regions
emerges by application of a vertically electric field in the
parallel configuration. In these energy ranges, one of the spin
types is fully blocked. Furthermore, the trace of energy band gap
is seen in Fig.\ref{fig7} where spin polarization is zero around
the zero energy.

{\it Magnetoresistance}: An in-plane electric field can change
the exchange field direction inducing in graphene. So by switching
between parallel and antiparallel configurations, a giant
magnetoresistance is achievable. Magnetic resistance is defined
as the following: \beq MR=1-\frac{G_{AP}}{G_{P}} \eeq where
$G^{p}=G_{up}^p+G_{down}^p$ and
$G^{ap}=G_{up}^{ap}+G_{down}^{ap}$ are conductance for parallel
and antiparallel configurations. Fig.\ref{fig8} shows
magnetoresistance as a function of Fermi energy. As we expect,
magnetoresistance reaches to its maximum value in the energy
window of $[-M,M]$. This large magnetoresistance is originated
from the energy gap produced in the antiparallel case. Therefore
such a system is a good candidate for spin valve devices.

In the last part of this work, let us have a comparison between
the parity selection rules in {\it monolayer} and {\it bilayer}
graphene nanoribbons. The parity conservation controls the
electronic transport properties in zigzag {\it monolayer}
graphene nanoribbons\cite{cheraghchi}. It is interesting to check
the validity of the parity conservation resulting from the
reflection symmetry in bilayer graphene.
\section{Violation of reflection symmetry in bilayer graphene nanoribbons}\label{Sec:ref-symmetry}
\begin{figure}
\centering
\includegraphics[width=10 cm,height=10cm]{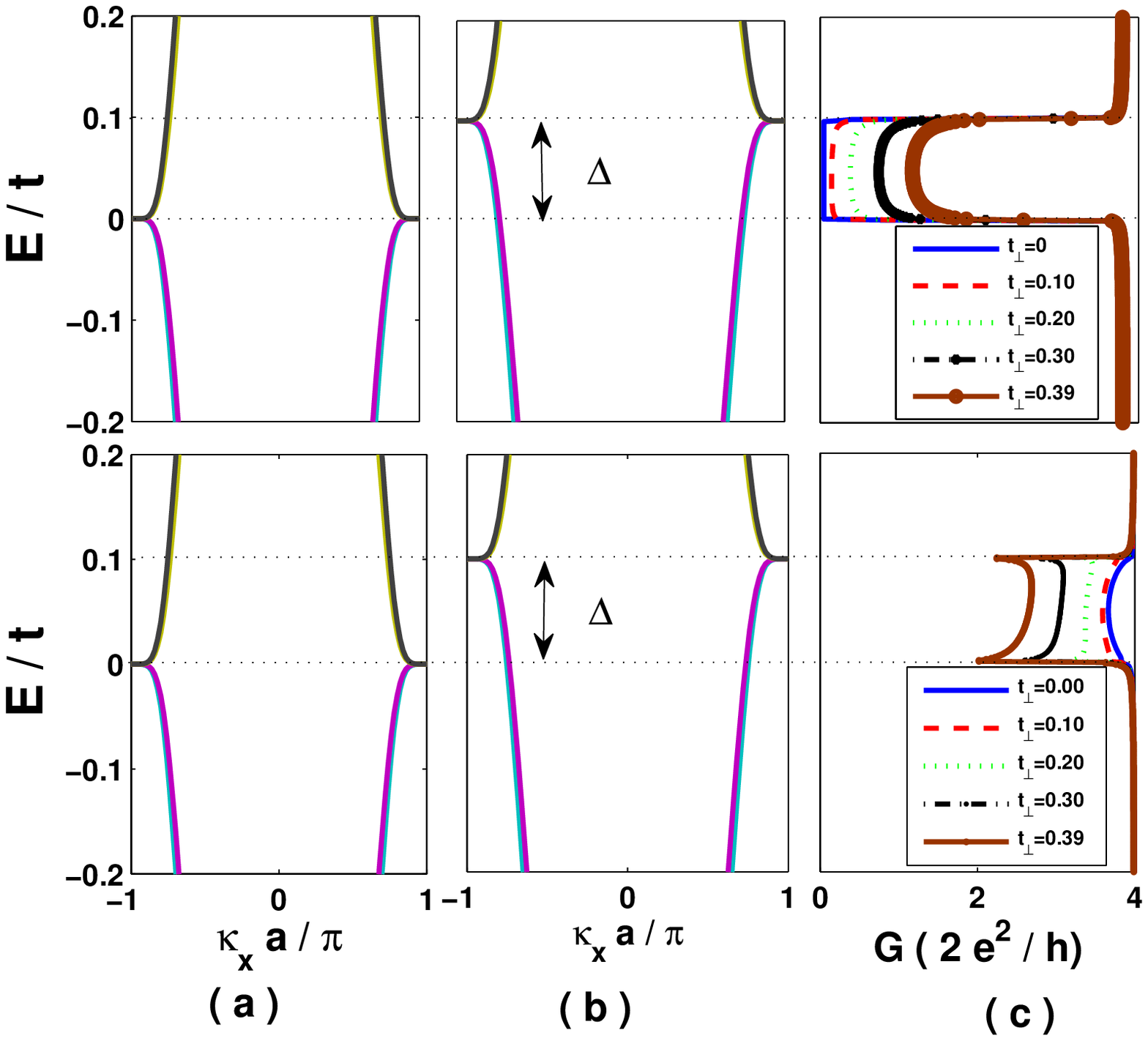}
\caption{c) Conductance through a step-like potential of zigzag
bilayer graphene nanoribbon with $\Delta=0.1t$ for $x>0$ and
$\Delta=0$ for $x<0$. Topper(Lower) curve is related to even
(odd) zigzag bilayer graphene nanoribbons with $6(5)$ zigzag
chains in width when intera-layer interaction ($t_{\perp}$) is
gradually increases. To present initial and final states for
scattering process, band structure of two regions is plotted in
parts (a) and (b). The left band structure belongs to the region
with zero potential while the right band structure belongs to the
region with the potential of $0.1t$. Energies are scaled by
$t$.}\label{fig2}
\end{figure}

Attaching the second layer to the monolayer graphene causes to
interact carbon atoms in the upper and lower layers with each
other. The intera-layer hopping integral energy between carbon
atoms located in different layers ($t_{\perp}$) is about $10$
times weaker than the inter-layer hopping energy ($t$). However,
nonzero $ t_{\perp}$ results in the violation of the reflection
symmetry in bilayer graphene nanoribbon when the number of zigzag
chains is even. To check this violation, we calculate conductance
through a step-like potential applied on ZBGN in which for $x<0$,
potential is zero and for $x>0$, gated potential is
$\Delta=0.1t$. Here, $x=0$ is located in the junction between the
left electrode and central region shown as in Fig.\ref{fig1}. In
this section, there is no applied perpendicular electric field
nor FM strips. Fig.\ref{fig2} shows conductance through a
step-like potential for even and odd ZBGN accompanied with the
left and right band structures for different perpendicular
hopping energies ($t_{\perp}$). By application of the gate
potential, band structures come out of the alignment such that
the upper bands are aligned with the lower bands. It is
interesting that if the connection between the upper and lower
layers is disconnected ($t_{\perp}=0$), parity conservation in
even ZGNRs is the only selection rule which governs electronic
transport\cite{cheraghchi} in the energy window of $0.1t$. In
this energy range and in the limit of two isolated monolayer
graphene, states belonging to the left side of junction have
opposite parity against states locating in the right side of the
junction. Therefore as shown in Fig.\ref{fig2}a, transport is
blocked. However, if interaction between the upper and lower
layers is gradually being turned on ($t_{\perp} \neq 0$), the
reflection symmetry is gradually being violated and the
electronic states located in the left side of the junction are
able to scatter into the states located in the right side of the
junction at the same energy. So conductance becomes nonzero in
the energy window of $[0,0.1t]$.

It is known that there is no reflection symmetry and thus parity
conservation in odd monolayer ZGNRs\cite{cheraghchi}. So in the
limit of two isolated graphene sheets ($t_{\perp}=0$), as shown in
Fig.\ref{fig2}b, transmission is complete through the step-like
potential in the energy window of [0,0.1$t$]. By increasing
interaction between two layers, intera-layer back-scattering
causes to lower transmission. As a result, when hopping between
atoms located in the upper and lower layers ($t_{\perp}$) reaches
to its realistic value $0.39 eV$, Fig.\ref{fig2} indicates that
there is no much difference in transport properties through even
or odd ZGNRs.
\section{Conclusion}\label{Sec:conclusion}
We study coherent spin dependent conductance through zigzag
bilayer graphene nanoribbon (ZBGN) in the presence of two
ferromagnetic insulator strips which are deposited on the edges
of nanoribbon. This system sets as normal/ferromagnetic/normal
graphene nanojunction. It is supposed that the induced exchange
potential decreases exponentially when one goes from the edges
into the middle of ribbon. Configuration of the exchange field
inducing by the edge ferromagnetic insulators can be set as
parallel or antiparallel with each other. A band gap is opened in
bilayer graphene when the exchange fields arising from two
ferromagnetic insulators are in the antiparallel situation. So in
this case, ZBGN behaves as a semiconductor. However, in the
parallel configuration of magnetization of strips, there is no
gap. In the parallel configuration, conductance indicates spin
polarized behavior around the zero energy. Spin polarization is
intensified by application of a perpendicular electric field. On
the other word, fully spin filter emerges in the energy ranges
which are equal to the induced exchange field. In addition, a
giant magnetoresistance is observed around the zero energy which
is proportional to the induced exchange potential. To be
comparable our results with the one for monolayer ZGNR, we
demonstrate that the reflection symmetry and so the parity
conservation fails in ZBGN.
\section{Acknowledgement}
All authors are grateful to Seyyed Ahmad Ketabi for useful
discussions at the early stages of the work. H. CH. thanks the
Condensed Matter Section of ICTP in Trieste for the hospitality
and support during his visit to this institute.

\end{document}